\documentclass[11pt,aps,prd,preprint,superscriptaddress]{revtex4-1}
\usepackage{amsmath}
\usepackage[dvips]{graphicx}
\usepackage[english]{babel}
\newcommand{\be}{\begin{equation}}
\newcommand{\ee}{\end{equation}}
\newcommand{\ben}{\begin{eqnarray}}
\newcommand{\een}{\end{eqnarray}}

\begin{document}
\title{Entropy evolution of universes with initial and final de Sitter eras}
\author{Jos\'{e} Pedro Mimoso \footnote{E-mail: jpmimoso@cii.fc.ul.pt}}
\affiliation{Faculdade de Ci\^{e}ncias e Centro de Astronomia e
Astrof\'{\i}sica da Universidade de Lisboa, 1749-016 Lisboa,
Portugal}
\author{Diego Pav\'{o}n\footnote{E-mail: diego.pavon@uab.es}}
\affiliation{Departamento de F\'{\i}sica, Universidad Aut\'{o}noma
de Barcelona, 08193 Bellaterra (Barcelona), Spain.}
\begin{abstract}
This brief report studies the behavior of entropy in two recent
models of cosmic evolution by J. A. S. Lima, S. Basilakos, and F.
E. M. Costa [Phys. Rev. D \underline{86}, 103534 (2012)], and J.
A. S. Lima, S. Basilakos, and J. Sol\'{a} [arXiv:1209.2802]. Both
start with an initial de Sitter expansion, go through the
conventional radiation and matter dominated eras to be followed by
a final and everlasting de Sitter expansion. In spite of their
outward similarities (from the observational viewpoint they are
arbitrary close to the conventional Lambda cold dark matter
model), they deeply differ in the physics behind them. Our study
reveals that in both cases the Universe approaches thermodynamic
equilibrium in the last de Sitter era in the sense that the
entropy of the apparent horizon plus that of matter and radiation
inside it increases and is concave. Accordingly, they are
consistent with thermodynamics. Cosmological models that do not
approach equilibrium at the last phase of their evolution appear
in conflict with the second law of thermodynamics.
\end{abstract}
\maketitle

\section{Introduction}
\noindent As daily experience teaches us, macroscopic systems tend
spontaneously to thermodynamic equilibrium. This constitutes the
empirical basis of the second law of thermodynamics. The latter
succinctly formalizes this  by establishing that the entropy, $S$,
of isolated systems never decreases, $S' \geq 0$, and that it is
concave, $S'' < 0$, at least in the last leg of approaching
equilibrium (see, e.g., \cite{callen}). The prime means derivative
with respect the relevant variable. In our view, there is no
apparent reason why this should not be applied to cosmic
expansion. Before going any further we wish to remark that
sometimes the second law is found formulated by stating just the
above condition on $S'$ but not on $S''$. While this  incomplete
version of the law works well for many practical purposes it is
not sufficient in general. Otherwise, one would witness systems
with an always increasing entropy but never achieving equilibrium,
something at stark variance with experience.
\  \\

\noindent In this paper we shall explore whether two recently
proposed cosmological models, that start from an initial de Sitter
expansion and  go through the conventional radiation and matter
eras to finally enter a never-ending de Sitter phase, present the
right thermodynamic evolution of above. This is to say, we will
see whether $S' \geq 0$ at all times and $S'' < 0$ at the
transition from the matter era to the final de Sitter one. By $S$
we mean the entropy  of the apparent horizon, $S_{h} = k_{B}\,
{\cal{A}}/ (4\,\ell_{pl}^{2})$, plus the entropy of the radiation,
$S_{\gamma}$, and/or pressureless matter, $S_{m}$, inside it. As
usual, ${\cal A}$ and $ \ell_{pl}$  denote the area of the horizon
and Planck's length, respectively.
\  \\

\noindent We shall consider the evolution of the entropy, first in
the model of Lima, Basilakos and Costa \cite{ademir2012} and then
in the model of Lima, Basilakos and Sol\'{a} \cite{jsola1} (models
I and II, respectively). Both assume a spatially flat
Friedmann-Robertson-Walker metric, avoid -by construction- the
horizon problem and the initial singularity of the big bang
cosmology, evolve between an initial and a final de Sitter
expansions (the latter being everlasting), and from the
observational viewpoint they are very close to the conventional
Lambda cold dark matter model.
\  \\

\noindent In spite of these marked coincidences the physics behind
the models is deeply different. While model I rests on the
production of particles induced by the gravitational field (and
dispenses  altogether with dark energy), model II assumes dark
energy in the form of a cosmological constant that in reality
varies with the Hubble factor in a  manner prescribed by quantum
field theory.
\  \\

\noindent We shall  focus on the transitions from the initial de
Sitter expansion to the radiation dominated era, and from the
matter era to the final de Sitter expansion. As is well known, in
the radiation and matter eras, as well as in the transition from
one to another, both $S'$ and $S''$ are positive-definite
quantities \cite{grg_nd1}.
\  \\

\noindent As usual, a zero subscript attached to any quantity
indicates that it is to be evaluated at the present time.

\section{Thermodynamic analysis of model I}
\noindent In model I the present state of cosmic acceleration is
achieved not by dark energy or as a result of modified gravity,
but simply by the gravitationally induced production of particles
\cite{ademir2012}. In this scenario the initial phase is a de
Sitter expansion which, due to the creation of massless particles
becomes unstable whence the Universe enters the conventional
radiation dominated era. At this point -as demanded by conformal
invariance \cite{parker1}- the production of these particles
ceases whereby the radiation becomes subdominant and the Universe
enters a stage dominated by pressureless matter (baryons and cold
dark matter). Lastly, the negative creation pressure associated to
the production of matter particles accelerates the expansion and
ushers the Universe in a never-ending de Sitter era. The model is
consistent with the observational tests, including the growth rate
of cosmic structures \cite{ademir2012}.

\subsection{From de Sitter to radiation dominated expansion}
\noindent The production of massless particles in the first de
Sitter era induces a negative creation pressure related to the
phenomenological rate of particle production, $\Gamma_{r}$, given
by $p_{c}= -(1+w) \rho \Gamma_{r}/(3H)$ -see \cite{ilya} for more
general treatments of the subject. Here, $\rho$ is the energy
density of the fluid (radiation in this case), $w$ its equation of
state parameter, and $H = \dot{a}/a$ the Hubble expansion rate. As
a consequence, the evolution of the latter is governed by
\begin{equation}
\dot{H} \, + \, \frac{3}{2}\, (1+w) H^{2} \, \left(1 \, - \,
\frac{\Gamma_{r}}{3H}\right) = 0  , \label{H1}
\end{equation}
\noindent cf. equation (7) in \cite{ademir2012}.

\noindent Because the rate must strongly decline when the Universe
enters the radiation dominated era, it may be modeled as $ \Gamma
_{r}/(3H) = H/H_{I}$ with  $H \leq H_{I}$, being $H_{I}$ the
initial de Sitter expansion rate. In consequence, for $w = 1/3$
(thermal radiation) last equation reduces to
\begin{equation}
\dot{H} \, + \, 2 H^{2} \left(1 \, - \, \frac{H}{H_{I}} \right) =
0 \, ,
\label{H2}
\end{equation}
\noindent whose solution in terms of the scale factor reads
\begin{equation}
H(a) = \frac{H_{I}}{1\, + \, D a^{2}} \,
\label{H3}
\end{equation}
\noindent with $D$ a positive-definite integration constant.
\  \\

\noindent The area of the apparent horizon  ${\cal A} = 4\pi
\tilde{r}^{2}_{\cal A}$, where $ \tilde{r}_{{\cal A}} =
\frac{1}{\sqrt{H^{2} + ka^{-2}}}\,$ is the radius \cite{bak-rey},
trivially reduces, in the case under consideration (a spatially
flat universe), to the Hubble length, $H^{-1}$. Accordingly, the
entropy of the apparent horizon, $S_{h} = k_{B} \pi /(\ell_{pl} \,
H)^{2}$, as the Universe transits  from de Sitter, $H = H_{I}$, to
a radiation dominated expansion is simply
\begin{equation}
S_{h} = \pi k_{B} \, \frac{(1\, +\, D a^{2})^{2}}{(\ell_{pl} \,
H_{I})^{2}} \, . \label{shor2}
\end{equation}
\noindent It is readily seen that $S_{h}$ is a growing, $S'_{h}
>0$, and convex, $S''_{h} > 0$, function of the scale factor (the
prime stands for $d/da$).

\noindent In its turn, the evolution of the entropy of the
radiation fluid inside the horizon  can be determined with the
help of Gibbs equation \cite{callen}
\begin{equation}
T_{\gamma}\, dS_{\gamma} = d\left(\rho_{\gamma}\, \frac{4 \pi}{3}
\tilde{r}^{3}_{{\cal A}}\right) \, + \, p_{\gamma} \, d
\left(\frac{4 \pi}{3} \, \tilde{r}^{3}_{{\cal A}}\right) \, ,
\label{gibbs1}
\end{equation}
\noindent where
\begin{equation}
\rho_{\gamma} = \rho_{I} \left[1 \, + \, \lambda^{2}
\left(\frac{a}{a_{*}} \right)^{2} \right]^{-2} \,
,
\label{rhogamma}
\end{equation}
\noindent $\rho_{I}$ is the critical energy density of the initial
de Sitter phase, $\lambda^{2} = D a_{*}^{2}$, and $p_{\gamma} =
\rho_{\gamma}/3$. In its turn, $a_{*}$ denotes the scale factor at
the transition from de Sitter to the beginning of the standard
radiation epoch.
\  \\

\noindent Likewise, the dependence of the radiation temperature on
the scale factor is given by
\begin{equation}
T_{\gamma}= T_{I} \left[1 \, + \, \lambda^{2}
\left(\frac{a}{a_{*}} \right)^{2} \right]^{-1/2}
\label{Tgamma}
\end{equation}
\noindent -cfr. Eq. (13) in \cite{ademir2012}.
\  \\

 \noindent In consequence,
\begin{equation}
T_{\gamma} \, S'_{\gamma} = \frac{4\pi}{3}\, \frac{\rho_{I}\,
D}{H_{I}^{3}}\, a > 0 \, . \label{gibbs2}
\end{equation}
\noindent Accordingly, $S'_{h} \, + \, S'_{\gamma} \geq 0$, i.e.,
the total entropy -which encompasses the horizon entropy plus the
entropy of the fluid in contact with it- does not decrease. In
other words, the generalized second law (GSL), first formulated
for black holes and their environment \cite{jakob} and later on
extended to the case of cosmic horizons \cite{extended}, is
satisfied.
\  \\

\noindent Let us now discern the sign of $S''_{h} \, + \,
S''_{\gamma}$. As we have already seen, $S''_{h} > 0$. As for
$S''_{\gamma}$, we insert $T_{\gamma} = T_{I} \, (1\, + \,
Da^{2})^{-1/2}$ into Eq.(\ref{gibbs2}) and obtain
\begin{equation}
S'_{\gamma} = \frac{4 \pi}{3} \, \frac{\rho_{I}\, D}{H^{3}_{I}\,
T_{I}}\,a (1 \, + \, D a^{2})^{1/2} \, .
\label{Sprimegamma}
\end{equation}
\noindent Thus,
\begin{equation}
S''_{\gamma} = \frac{4 \pi}{3} \, \frac{\rho_{I} D}{H^{3}_{I}\,
T_{I}} \, \left[\frac{1 \, + \, 2D\, a^{2}}{(1 \, + \, D \,
a^{2})^{1/2}}\right] > 0 \, .
\label{sgammapprime}
\end{equation}
\noindent Therefore,  $S''_{h} \, + \, S''_{\gamma} > 0$; that is
to say, in the transition  from the initial de Sitter expansion to
radiation domination, the total entropy is a convex function of
the scale factor. If it were concave, the Universe could have
attained a state of thermodynamic equilibrium and would have not
left it unless forced  by some ``external agent". The initial de
Sitter expansion ($H = H_{I}$, and no particles) was a state of
equilibrium, but only a metastable one for the Universe was
obliged to leave it by the production of particles which acted as
an external agent.

\subsection{From matter domination to the final de Sitter expansion}
\noindent The Hubble function of spatially flat Lambda cold dark
matter models obeys
\begin{equation}
\dot{H} \, + \, \frac{3}{2} H^{2} \, \left[1\, - \,
\left(\frac{H_{\infty}}{H}\right)^{2} \right] = 0 \, ,
\label{dotH1}
\end{equation}
\noindent where $H_{\infty} = \,$ constant denotes its asymptotic
value at the far future.
\  \\

\noindent In a dust filled universe ($w = 0$) with production rate
$\Gamma_{dm} \leq 3H$ of pressureless matter the Hubble factor
obeys
\begin{equation}
\dot{H} \, + \, \frac{3}{2}\, H^{2}\, \left(1 \, - \,
\frac{\Gamma_{dm}}{3H} \right) = 0 \, .\label{dotH2}
\end{equation}
\noindent Comparison with the previous equation leads to
$\Gamma_{dm}/(3H) = (H_{\infty}/H)^{2}\, $, i.e., $\Gamma_{dm}
\propto H^{-1}$.
\  \\

\noindent Thus,
\begin{equation}
H^{2} = H_{0}^{2} \, [\tilde{\Omega}_{m} \, a^{-3} \, + \,
\tilde{\Omega}_{\Lambda}] \, ,
\label{Hsquare}
\end{equation}
\noindent where $\tilde{\Omega}_{\Lambda} = (H_{\infty}/H_{0})^{2}
= 1 \, - \, \tilde{\Omega}_{m} = \,$ constant $ >0$.
\  \\

\noindent Recalling that $S_{h} = k_{B}\,  {\cal{A}}/
(4\,\ell_{pl}^{2})$, it follows $S'_{h} = -2 \pi k_{B} \,
H'/(\ell_{pl}^{2} \, H^{3})$ with
\begin{equation}
H' = - \frac{3}{2}\, H_{0} \, \frac{(1 \, - \,
\tilde{\Omega}_{\Lambda})\, a^{-4}}{[(1 \, - \,
\tilde{\Omega}_{\Lambda})\, a^{-3}\, + \, \tilde{\Omega}_{\Lambda}
]^{1/2}} \, .
\label{Hprime}
\end{equation}
\noindent Then
\begin{equation}
S'_{h} = 6 \frac{\pi k_{B}}{\ell_{pl}^{2}} \, \frac{(1 \, - \,
\tilde{\Omega}_{\Lambda})\, a^{-4}}{H_{0}^{2}\, [(1 \, - \,
\tilde{\Omega}_{\Lambda})\, a^{-3}\, + \, \tilde{\Omega}_{\Lambda}
]^{2}} > 0 \, . \label{shorprime2}
\end{equation}
\  \\

\noindent As for the entropy of dust matter, it suffices to
realize that every single particle contributes to the entropy
inside the horizon by a constant bit, say $k_{B}$. Then, $S_{m} =
k_{B} \frac{4 \pi}{3} \tilde{r}^{3}_{{\cal A}}\, n$, where  the
number density of dust particles obeys the conservation equation $
n' = (n/(aH)) [\Gamma_{dm} \, - \, 3H] < 0 $ with $\Gamma_{dm} =
3H_{0}^{2}\,\tilde{\Omega}_{\Lambda}/H > 0$.
\  \\

\noindent Thus,
\begin{equation}
S'_{m} = \frac{4 \pi}{3} k_{B}\, \frac{n}{H^{4}}
\left[\frac{\Gamma_{dm}\, - \, 3H}{a} \, - \, 3 H'\right] \, .
\label{smprime}
\end{equation}
\noindent Since $\Gamma_{dm} -  3H < 0 $ and $H' < 0$ the sign of
$S'_{m}$ is undecided at this stage. To ascertain it consider the
square parenthesis in (\ref{smprime}) and multiply it by $aH/3$.
One obtains
\begin{equation}
\frac{aH}{3}\, \left[\frac{\Gamma_{dm}\, - \, 3H}{a} \, - \, 3
H'\right] = \textstyle{1\over{2}} H_{0}^{2}\, (1\, - \,
\tilde{\Omega}_{\Lambda})\, a^{-3} > 0. \label{oneobtains}
\end{equation}
\noindent In consequence, $S'_{m} > 0$ and the GSL, $S'_{h} \, +
\, S'_{m} \geq 0$, is satisfied also in this case.
\  \\

\noindent Let us now consider the sign of $S''_{h} \, + \,
S''_{m}$ in the limit $a \rightarrow \infty$. From $S'_{h} = -
\frac{2 \pi k_{B}}{\ell_{pl}^{2}} \, (H'/H^{3})\, $ it follows,
\begin{equation}
S''_{h} = - \frac{2 \pi k_{B}}{\ell_{pl}^{2}} \frac{1}{H^{4}}\, [H
H'' \, - \, 3H'^{2} ]. \label{Shorpprime2}
\end{equation}
\noindent In virtue of (\ref{Hprime}) we get
\begin{equation}
H \, H'' = \frac{3}{2}H_{0}^{2}\, \left\{
\frac{4(1-\tilde{\Omega}_{\Lambda})[(1-\tilde{\Omega}_{\Lambda})a^{-3}+
\tilde{\Omega}_{\Lambda}]a^{-5} \, + \,
(3/2)(1-\tilde{\Omega}_{\Lambda})^{2} \,
a^{-8}}{(1-\tilde{\Omega}_{\Lambda})a^{-3} \, + \,
\tilde{\Omega}_{\Lambda}} \right\} \, , \label{hhpprime}
\end{equation}
\noindent whence
\begin{equation}
HH'' \, - \, 3H'^{2} = \frac{3}{2}\, H_{0}^{2} \,
\left\{\frac{4(1-\tilde{\Omega}_{\Lambda})[(1-\tilde{\Omega}_{\Lambda})a^{-3}+
\tilde{\Omega}_{\Lambda}]a^{-5}}{(1-\tilde{\Omega}_{\Lambda})a^{-3}
\, + \, \tilde{\Omega}_{\Lambda}}\right\} > 0 \, .
\label{hhpprime}
\end{equation}
\noindent Thereby, in view of  (\ref{Shorpprime2}) we get $S''_{h}
<0$.
\  \\

\noindent As for the sign of $S''_{m}$ it suffices to recall, on
the one hand,  Eq. (\ref{smprime}) and realize that $\Gamma_{dm}(a
\rightarrow \infty) =
3H_{0}^{2}\tilde{\Omega}_{\Lambda}/H_{\infty} = 3H_{\infty}$ and
that $H'(a \rightarrow \infty) \rightarrow 0$; then, $S'_{m}(a
\rightarrow \infty) = 0$. And, on the other hand, that $S'_{m} (a
< \infty) > 0$. Taken together they imply that $S'_{m}$ tends to
zero from below, i.e., that $S''_{m}(a \rightarrow \infty) <0$.
\  \\

\noindent Altogether,  when $a \rightarrow \infty$ one has
$S''_{h}\, + \, S''_{m} <0$, as expected. Put another way, in the
phenomenological model of Lima {\it et al.} \cite{ademir2012} the
Universe behaves as an ordinary macroscopic system \cite{grg_nd2};
i.e., it eventually tends to thermodynamic equilibrium, in this
case characterized by a never-ending de Sitter expansion era with
$H_{\infty} = H_{0}\, \sqrt{\tilde{\Omega}_{\Lambda}} < H_{0}$.

\section{Thermodynamic analysis of model II}
\noindent The model of Ref. \cite{jsola1} is based on the
assumption that in quantum field theory in curved spacetime the
cosmological constant is a parameter  that runs with the Hubble
rate in a specified manner \cite{parker2,jsola2}. As in the
previous model, the vacuum decays into radiation and
nonrelativistic particles while the Universe expands from de
Sitter to de Sitter through the intermediate eras of radiation and
matter domination. However, as said above, the physics of both
models differ drastically from one another. While model I
dispenses altogether with dark energy, model II assumes dark
energy in the form of cosmological term, $\Lambda$, that evolves
with expansion.
\  \\

\noindent According  to model II, the running of $\Lambda$ is
given by the sum of even powers of the Hubble expansion rate,
\begin{equation}
\Lambda(H) = c_{0} \, + \, 3 \nu H^{2} + \, 3 \alpha
\frac{H^{4}}{H_{I}^{2}} \, ,
\label{Lambda(H)1}
\end{equation}
\noindent where $c_{0}$, $\alpha$ and $\nu$ are constants of the
model. The absolute value of the latter  is constrained by
observation as $|\nu| \sim 10^{-3}$. At early times the last term
dominates, and at late times ($H \ll H_{I}$) it becomes negligible
whereby (\ref{Lambda(H)1}) reduces to
\begin{equation}
\Lambda(H) =  \Lambda_{0} \, +\, 3 \nu (H^{2}\, - \, H_{0}^{2}) \,
\label{Lambda(H)2}
\end{equation}
\noindent with $\Lambda_{0} = c_{0} \, + \, 3 \nu H_{0}^{2}$.
\  \\

\noindent At the early universe, integration of the field
equations gives for the Hubble function, the energy density of
radiation and of vacuum, the following  expressions \cite{jsola1}
\begin{equation}
H(a) = \sqrt{\frac{1-\nu}{\alpha}} \, \frac{H_{I}}{\sqrt{D\, a^{3
\beta}+1}}  \qquad \; \; (\beta = (1-\nu)(1+w))\, ,
\label{H(a)running1}
\end{equation}
\begin{equation}
\rho_{\gamma} = \rho_{I} \frac{(1-\nu)^{2}}{\alpha} \frac{D\, a^{3
\beta}}{[D\, a^{3 \beta}\, +\, 1]^{2}}\, , \; \; {\rm and} \; \;
\rho_{\Lambda} = \frac{\Lambda}{8 \pi G} = \rho_{I} \frac{1 -
\nu}{\alpha} \, \frac{\nu \, D\, a^{3 \beta}\, +\, 1}{[D\, a^{3
\beta}\, + \, 1]^{2}}\, ,\label{rhorunning1}
\end{equation}
where $D (>0)$ is an integration constant.
\  \\

\noindent As in model I, at the transition from de Sitter to the
radiation era, the first and second derivatives of $S_{h}$ and
$S_{\gamma}$ (with respect to the scale factor) are all positive.
Thus, at this transition the GSL is fulfilled but neither the
radiation era nor the subsequent matter era correspond to
equilibrium states since at them $S'' > 0$.
\  \\

\noindent By integration of the field equations at late times it
is seen that  the transition between the stages of matter
domination to the second (and final) de Sitter expansion is
characterized by
\begin{equation}
H(a) = \frac{H_{0}}{\sqrt{1-\nu}}\, \sqrt{(1-\Omega_{\Lambda 0})\,
a^{-3(1-\nu)}\, + \, \Omega_{\Lambda 0}- \nu}\, ,
\label{H(a)running2}
\end{equation}
\begin{equation}
\rho_{m} = \rho_{m 0}\,  a^{-3(1-\nu)}\, \quad {\rm and} \quad
\rho_{\Lambda}(a) = \rho_{\Lambda 0} \, + \, \frac{\nu}{1-\nu}\,
\rho_{m0} \, \left[a^{-3 (1-\nu)}\, - \, 1  \right]\, ,
\label{rhorunning2}
\end{equation}
where $\Omega_{\Lambda 0}=  8\pi G \rho_{\Lambda
0}(3H_{0}^{2})^{-1}$. From the pair of equations
(\ref{rhorunning2}) we learn that dust particles are created out
of the vacuum at the rate $\Gamma_{dm} = \nu H$.
\  \\

\noindent As in the previous model, $S_{h}' >0$ and $S_{h}'' <0$.
However at variance with it, the matter entropy, $S_{m} = k_{B}
\frac{4 \pi}{3} \tilde{r}^{3}_{{\cal A}}\, n \propto H^{-3}\, n$,
decreases with expansion and is convex. This is so because, in
this case,  the rate of particle production, $\Gamma_{dm}$, goes
down and  cannot compensate for the rate of dilution caused by
cosmic expansion. Nevertheless, as it can be easily checked,
$S'_{h}$ and $S''_{h}$ dominate over $S'_{m}$ and $S''_{m}$,
respectively, as $a \rightarrow \infty$. Thus,  as in  model I,
the total entropy results a growing and concave function of the
scale factor, at least at the far future stage. Hence, the
Universe gets asymptotically closer and closer to thermodynamic
equilibrium.

\section{Discussion and concluding remarks}
\noindent The second law of thermodynamics constrains the
evolution of macroscopic systems; thus far, all attempts to
disprove it by means of ``counterexamples" have failed. While it
seems reasonable to expect it to be obeyed also by the Universe as
a whole, a proof of this on first principles is still lacking.
However, persuasive arguments based on the Hubble history,
suggesting that, indeed, the Universe behaves as any ordinary
macroscopic thermodynamic system (i.e., that it tends to a maximum
entropy state), were recently given in \cite{grg_nd2}. On the
other hand, in view of the strong connection between  gravitation
and thermodynamics -see e.g., \cite{tedj, padm}- it would be
shocking that the Universe behaved otherwise. In this spirit we
have considered models \cite{ademir2012} and \cite{jsola1}, each
of them covering the whole cosmic evolution (i.e., the two
well-known eras of radiation and matter dominance sandwiched
between an initial and a final de Sitter expansions), and
consistent with recent observational data. In both models, the
entropy, as a function of the scale factor, never decreases and is
concave at least at the last stage of evolution, signaling that
the Universe is finally approaching thermodynamic equilibrium.
\  \\

\noindent In principle, the initial de Sitter eras should be
stable ($H$ and $S$ are constants when $t \rightarrow - \infty$)
but owing to particle production,  which can be viewed as
``external" agent acting on the otherwise isolated system, an
instability sets in. Once the Universe gets separated from
thermodynamic equilibrium it reacts trying to restore it -as
ordinary systems do-, only that at lower energy scale. This is
finally achieved at the last de Sitter expansion. Given that two
de Sitter expansions cannot directly follow one another  an
intermediate phase (comprised by the radiation and matter eras) is
necessary in between.
\\

\noindent As is well known, irreversible particle production, as
is the case in models I and II, implies generation of entropy (see
e.g., \cite{ilya}, \cite{ademir1996}); something rather natural
because the new born particles necessarily increase the volume of
the phase space. Our analysis takes this into account in an
implicit and straightforward manner via the $\Gamma$ rates of
particle production. These quantities modify the corresponding
expressions for the Hubble factor and hence $S'$ and $S''$. For
instance, setting $\Gamma_{r}$ to zero in Eq. (\ref{H1}) (which
would kill model I) leads to $D = 0$ and therefore to $S'_{\gamma}
= S''_{\gamma} = 0$ (Eqs. (\ref{Sprimegamma}) and
(\ref{sgammapprime}), respectively). Likewise if, in the same
model, one sets $\Gamma_{dm}$ to zero, then $S'_{m}$ (Eq.
(\ref{smprime})) decreases. Analogous statements can be made about
model II if the parameter $\nu$ (that enters the corresponding
$\Gamma$ rates) is forced to vanish (again, this would kill the
model).
\\

\noindent When quantum corrections to Bekenstein-Hawking entropy
law are taken into account,  the entropy of black hole horizons
generalizes to $ S_{h} = k_{B} \, \left[
\frac{{\cal{A}}}{4\,\ell_{pl}^{2}} \, - \, \frac{1}{2} \, \ln
\left(\frac{{\cal{A}}}{\ell_{pl}^{2}}\right)\right]$ plus higher
order terms \cite{meissner,ghosh}. Assuming this also applies to
the cosmic apparent horizon, one may wonder up to what extent this
may modify our findings. The answer is that the modifications are
negligible whereby our results are robust  against  quantum
corrections to the Bekenstein-Hawking entropy.  We illustrate this
point by noting  that the expression for $S'_{h}$ of model I in
the transition from the initial de Sitter regime to radiation
domination presents now the overall multiplying factor $\left\{1\,
- \, \frac{\ell_{pl}^{2}\, H_{I}^{2}}{8\pi
(1+Da^{2})^{2}}\right\}$. In this expression the second term is
negligible on account of the quantity $\ell_{pl}^{2}$ in the
numerator. It is noteworthy that the imposing of the condition
$S'_{h} >0$ leads to $\frac{\ell_{pl}^{2}\, H_{I}^{2}}{8 \pi
(1+Da^{2})^{2}} < 1$. When this inequality is evaluated in the
limit $a \rightarrow 0$,  the upper bound on the square of the
initial Hubble factor, $H_{I}^{2} < 8 \pi/\ell_{pl}^{2}$ follows.
Thus, the nice and convincing result, that the initial Hubble
factor cannot be arbitrarily large (its square, not much larger
than Planck's curvature) arises straightforwardly from the quantum
corrected Bekenstein-Hawking entropy law.
\\

\noindent Likewise, a study of $S'_{h}$ of model II in the
transition from de initial Sitter expansion to radiation
domination leads, in the same limit of very small $a$, to the
upper bound $ H_{I}^{2} < 2\pi \alpha/[(1-\nu) \ell_{pl}^{2}]$,
i.e., to essentially identical result on the maximum permissible
value of $H_{I}$. It is remarkable that in spite of being models I
and II so internally different, they share this bound.
\\

\noindent We conclude that models I and II show consistency with
thermodynamics,  and that their overall behavior (in particular,
the reason why they  evolve precisely to de Sitter in the long
run) can be most easily understood from the thermodynamic
perspective. Further, these results remain valid also if quantum
corrections to Bekenstein-Hawking entropy law are incorporated.
\  \\

\noindent It would be interesting to explore the possible
connection of the second law when applied to expanding universes
with the ``cosmic no-hair conjecture" \cite{gibbons}. Loosely
speaking, the latter asserts that ``all expanding-universe models
with positive cosmological constant asymptotically approach the de
Sitter solution" \cite{bob}. There is an ample body of literature
on this -see e.g., \cite{jdbarrow} and \cite{cotsakis} and
references therein. In the light of the above we may venture to
speculate that the said conjecture and the tendency to
thermodynamic equilibrium at late times are closely interrelated.
Nevertheless, this is by no means the last word as the subject
calls for further study.
\  \\

\noindent Before closing, note that the particle production does
not vanish in the long run ($\Gamma_{dm}(a \rightarrow \infty) = 3
H_{\infty}$, and $\Gamma_{dm}(a \rightarrow \infty) =
\frac{\nu}{1-\nu}H_{0}\sqrt{\Omega_{\Lambda 0} - \nu}$ in models I
and II, respectively). Then, the question arises as to whether it
will be strong enough to bring instability on the second (an in
principle, final) de Sitter expansion. Our tentative answer  is in
the negative; the reason being that in both cases the expansions
tend to strictly de Sitter ($H = {\rm constant} >0$). However, a
definitive response requires far more consideration and it lies
beyond the scope of this work. At any rate, if instability sets in
again, one should expect that the whole story repeats itself anew
though at a much lower energy.

\acknowledgments{D.P. is grateful to the ``Centro de Astronomia e
Astrof\'{\i}sica da Universidade de Lisboa", where part of this
work was done, for warm hospitality and financial support. This
research was partially funded by the Portuguese Science Foundation
(FCT) through the projects CERN/FP/123618/2011 and
CERN/FP/123615/2011, as well as PTDC/FIS/102742/2008, and the
``Ministerio Espa\~{n}ol de Educaci\'{o}n y Ciencia" under Grant
No. FIS2012-32099, and by the ``Direcci\'{o} de Recerca de la
Generalitat" under Grant No. 2009SGR-00164.}


\end{document}